\documentclass[onecolumn]{revtex4}
\usepackage{amsmath,amssymb,graphics,epsfig,subfigure}
\usepackage{color}
\usepackage[colorlinks,linkcolor=red,anchorcolor=red,citecolor=green]{hyperref}
\usepackage{setspace}
\usepackage{booktabs}
\usepackage{float}
\begin{document}
\thispagestyle{empty}

\begin{center}
\title{Thermodynamic phase transition and winding number for the third-order Lovelock black hole}

\author{Yu-Shan Wang, Zhen-Ming Xu\footnote{E-mail: zmxu@nwu.edu.cn}, and Bin Wu
        \vspace{6pt}\\}

\affiliation{$^{1}$School of Physics, Northwest University, Xi'an 710127, China\\
$^{2}$Shaanxi Key Laboratory for Theoretical Physics Frontiers, Xi'an 710127, China\\
$^{3}$Peng Huanwu Center for Fundamental Theory, Xi'an 710127, China}

\begin{abstract}
Phase transition is important for understanding the nature and evolution of the black hole thermodynamic system. In this study, the connection between the phase transition of a black hole and the winding number derived by the complex analysis is used to predict the type of the black hole phase transition. For the third-order Lovelock black holes, at the hyperbolic topology in any dimensions and the spherical topology in $7$ dimensions, we arrive at the winding numbers both are $W=3$ which predicts that the system will undergo both the first-order and second-order phase transitions. For the spherical topology in $7<d<12$ dimensions, the winding number is $W=4$ and the corresponding phase transition will occur in two situations: one with only pure second-order phase transition and the other with both first-order and second-order phase transitions. We further confirm the correctness and rationality of this prediction by placing the black hole thermodynamics system in the potential field.
\end{abstract}

\maketitle
\end{center}

\section{Introduction}\label{Intro}

Black hole is an extreme celestial body predicted by the general relativity~\cite{Curiel2018cbt}. Inspired by the presentation of the Bekenstein's entropy~\cite{Bekenstein1973ur} for the black hole, Hawking concluded that when the quantum effect is taken into account, a black hole emits thermal radiation just like a normal black body. This means that the black hole has a temperature. The idea that black holes possess entropy and temperature is undoubtedly one of the most important discoveries of the 20th century and has been a topic of discussion for decades.

A central element of black hole thermodynamics is the phase transition, i.e., the transition from one state to another, accompanied by abrupt changes in physical quantities such as energy, entropy and volume under different parameter conditions. Hawking and Page first investigated the thermodynamic properties of the Anti-de Sitter (AdS) black hole and found that there is a phase transition between the Schwarzschild AdS black hole and pure AdS thermal radiation, i.e., the famous Hawking-Page phase transition~\cite{Hawking1982dh}. Subsequently, the black hole thermodynamics ushered in groundbreaking achievements under the the pioneering work~\cite{Kastor2009wy}. The extended phase space of the AdS black hole thermodynamics was introduced, where the negative cosmological constant is considered as the effective thermodynamic pressure of the black hole and its conjugate quantity is the thermodynamic volume, which initiated the recent surge of interest in extended black hole thermodynamics. The small-large black hole phase transition presented by the charged AdS black hole thermodynamic system has a more direct and precise overlap with the van der Waals system~\cite{Chamblin1999hg,Dolan2011xt,Wei2019uqg,Wei2015iwa,Kubiznak2012wp,Niu2011tb,Bhattacharya2017nru}. Currently, the study of the phase transition of black holes in extended phase space has been widely applied to various complex scenarios~\cite{Mo2014qsa,Miao2016ulg,Guo2021wcf,Qu2022nrt,Guo2023pob,Guo2022cdj}.

In addition, the holographic thermodynamics~\cite{Ahmed2023dnh,Cong2021fnf,Visser2022,Gong2023ywu} and the restricted phase space thermodynamics~\cite{Kong2022gwu,Kong2022tgt,Gao2021xtt,Zeyuan2021uol} have been proposed to give a holographic interpretation of black hole thermodynamics and to make black hole thermodynamics more like ordinary thermodynamics. Moreover, the topology has emerged as a new way to describe the type of phase transition in black holes. In the study~\cite{Wei2021vdx,Wei2022dzw}, it is described in detail how to use the $\phi$-map topological flow theory to construct a topological number that is independent of the endogenous parameters of black holes. The topological number can be used to distinguish between locally stable and locally unstable black hole phases, as well as to topologically classify the same class of black holes~\cite{Yerra2022alz,Yerra2022coh,Wu2023sue,Fang2022rsb,Bai2022klw}. These studies can deepen our understanding of black hole physics and contribute to go for clues to reveal the nature of black holes and the quantum theory of gravity.

The analysis of the type and criticality of the thermodynamic phase transition in black holes currently dominates the investigations. The swallowtail diagrams of the Gibbs free energy can give certain answers about the macroscopic thermodynamic processes of black hole phase transitions, but overlook the details of phase transition. Some ideas have been proposed to use the free energy landscape~\cite{Li2020nsy,Yang2021ljn,Li2022oup} and Landau free energy~\cite{Xu2021qyw} to explore the evolutionary processes associated with black hole phase transitions.

In a recent study~\cite{Xu2021usl}, the author constructed a thermal potential to study the black hole phase transition. The thermal potential is
\begin{equation}\label{dU}
U=\int(T_h-T)dS,
\end{equation}
where $T_h$ is the Hawking temperature of the black hole, $T$ is the canonical ensemble temperature and $S$ is the entropy of the black hole. The $U$, $T_h$ and $S$ are the functions of the radius of the event horizon $r_h$ and $T$ is just a positive constant which can be assigned in any way. When a standard system is determined to be a black hole, then the ensemble temperature of the system should be  the Hawking temperature of the black hole, i.e., $T=T_h$. Similar to the fluctuation, the thermal potential shows that all possible other thermodynamic states of the system deviate from the black hole states. From Eq.~\eqref{dU}, it follows that the extremum of the potential represents all possible black hole states,
\begin{equation}
	\dfrac{dU}{dS}=0~~\Rightarrow~~T=T_h.
\end{equation}
More importantly, the concave (convex) nature of the thermal potential represents the stable (unstable) state of the black hole
\begin{equation}
\delta \Biggl( \dfrac{dU}{dS}\bigg|_{T=T_h} \Biggr)=\left\{
\begin{aligned}
	\dfrac{\partial T_h(r_h)}{\partial S(r_h)}\bigg|_{T=T_h}>0,&~~~~\rm{stable ~ case};\\
	\dfrac{\partial T_h(r_h)}{\partial S(r_h)}\bigg|_{T=T_h}<0,&~~~~\rm{unstable ~ case}.
\end{aligned}
\right.
\label{variation}
\end{equation}

\begin{figure}[htb]
\begin{center}
 \includegraphics[width=70 mm]{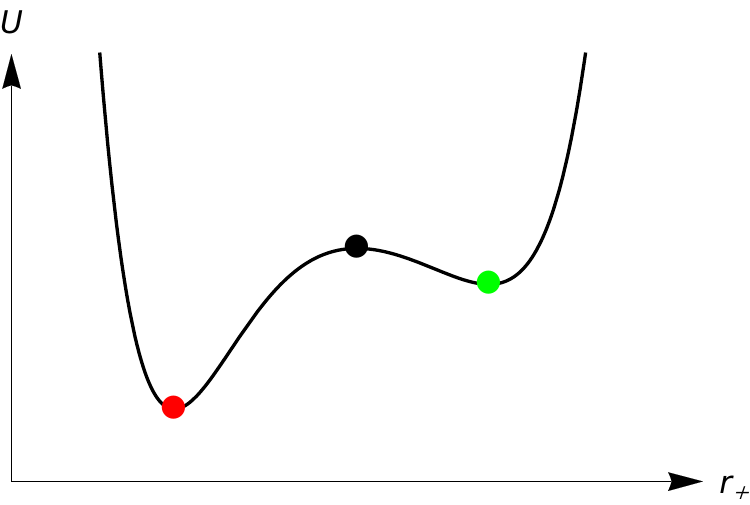}
\caption{The schematic diagram of thermal potential, where the \textcolor[rgb]{1.00,0.00,0.00}{$\bullet$} represents the global minimum, the $\bullet$ represents the local maximum and the--\textcolor[rgb]{0.00,1.00,0.00}{$\bullet$} represents the local minimum.}
\end{center}
\label{fig1}
\end{figure}

The schematic diagram of thermal potential described by Eq.~\eqref{dU} is shown in Fig.~\ref{fig1}. It is certain that the lowest point (the red point) is the most stable state in the entire canonical ensemble. As the different parameters of the black hole change, the extreme point of the thermal potential constantly changes, which corresponds to the changes between the black hole state and other unknown states in the ensemble. In the framework, we studied the microscopic phase transition mechanism of the charged AdS black holes~\cite{Xu2022jyp} and found that the phase transition of the large and small black holes exhibits severely asymmetric features, which fills the gap in the analysis of stochastic processes in the first-order phase transition rate problem of AdS black holes.

In four-dimensional spacetime, Einstein gravity can give the most appropriate explanation. While in higher dimensions, when the energy approaches the Planck energy scale, the high-order curvature terms of spacetime cannot be neglected and Einstein's general relativity theory requires some modifications. One of the widely accepted and valid candidates is the Lovelock gravity. Naturally, Lovelock gravity is an extension of Einstein gravity in higher dimensional spacetime and it proposes that the quantities acting in higher dimensional gravity should contain high-order gauge terms. The black hole solution in this gravity and the associated thermodynamic properties have been much studied~\cite{Cai2003kt,Kastor2010gq,Kastor2011qp,Khuzani2022lqd}. When we consider third-order Lovelock gravity, its action contains four terms: the cosmological constant term, the Einstein action term, the Gauss-Bonnet term, and the third-order Lovelock term. Black hole thermodynamics under third-order Lovelock gravity has also been much studied~\cite{Dehghani2005vh, Zou2010yr, Xu2014tja, Xu2016tja, Farhangkhah2021tzq}. Then the specific details behind its phase transition become the main object to study. Therefore, it inspires us to explore and analyse the microscopic processes of the phase transition of small and large black holes in third-order Lovelock black holes. Through the thermal potential and complex analysis, we study how a black hole transforms from one black hole state to another one under the influence of temperature $T$ and pressure $P$ in order to obtain its specific transition processes. We wish to further enrich the black hole phase transition dynamics process.

The structures of the paper are as follows. In Sec.~\ref{Review}, we give a brief introduction to third-order Lovelock black holes. Then, the winding number is related to black hole thermodynamics using a complex analysis approach. In Sec.~\ref{hyper}, the phase transition in the hyperbolic case is studied, focusing on $d = 7$. In Sec.~\ref{spher}, the phase transition in the spherical case is further studied, focusing on the analysis of $d = 7$ and $d=9$ cases. Finally, Sec.~\ref{Summary} is devoted to summary and discussion.

\section{Review of the third order Lovelock black Hole}\label{Review}

First, the $d$-dimensional Lovelock lagrangian density is~\cite{Myers1988ze,Aiello2004rz,Deruelle1989fj}
\begin{align}
		\mathcal{L}=&\sum_{n=0}^N \alpha_n \lambda^{2(n-1)} \mathcal{L}_n,
		\label{L}\\
		\mathcal{L}_n=&\frac{1}{2^n} \sqrt{-g} \delta_{j_1 \ldots j_{2 n}}^{i_1 \ldots i_{2 n}} R_{i_1 i_2}^{j_1 j_2} \ldots R_{i_{2 n-1} i_{2 n}}^{j_{2 n-1} j_{2 n}},
		\label{L_n}
	\end{align}	
	where
	\begin{equation}
		N=\left\{
		\begin{aligned}
			\dfrac{d}{2}-1,&~~~~ \rm{for ~even }~ d,\\
			\dfrac{d-1}{2},&~~~~\rm{for~ odd}~ d,
		\end{aligned}
		\right.\\
		\label{N}
	\end{equation}
	and $n$ is the order, $\alpha_n$ and $\lambda$ are the coupling constants for each of the lagrangian density functions,  $g$ is the determinant of the metric $g_{\mu\nu}$, $R^{\lambda}{}_{\mu \nu \rho}$ is the Riemann tensor , $R^{\lambda\mu}{}_{ \nu \rho}=g^{\mu \beta}R^{\lambda}{}_{\beta \nu \rho}$ and  $\delta_{j_1 \ldots j_{2 n}}^{i_1 \ldots i_{2 n}}$ is the generalized Kronecker delta of order $2n$. Of course, for calculation, here we try to list the first four items of the lagrangian,
	\begin{align}
		\mathcal{L}_0= & \sqrt{-g}, \\
		\mathcal{L}_1= & \frac{1}{2} \sqrt{-g} \delta_{j_1 j_2}^{i_1 i_2} R_{i_1 i_2}^{j_1 j_2}=\sqrt{-g} R,  \\
		\mathcal{L}_2= & \frac{1}{4} \sqrt{-g} \delta_{j_1 j_2 j_3 j_4}^{i_1 i_2 i_3 i_4} R_{i_1 i_2}^{j_1 j_2} R_{i_3 i_4}^{j_3 j_4}
		=\sqrt{-g}\left(R_{\mu \nu \rho \sigma} R^{\mu \nu \rho \sigma}-4 R_{\mu \nu} R^{\mu \nu}+R^2\right),\\		
		\mathcal{L}_3 = & \frac{1}{8} \sqrt{-g} \delta_{j_1 j_2 j_3 j_4 j_5 j_6}^{i_1 i_2 i_3 i_4 i_5 i_6} R_{i_1 i_2}^{j_1 j_2} R_{i_3 i_4}^{j_3 j_4} R_{i_5 i_6}^{j_5 j_6}  \\
		= & \sqrt{-g} (R^3 +2 R^{\mu \nu \sigma \kappa} R_{\sigma \kappa \rho \tau} R_{\mu \nu}^{\rho \tau}+8 R_{\sigma \rho}^{\mu \nu} R_{\nu \tau}^{\sigma \kappa} R^{\rho \tau}+24 R^{\mu \nu \sigma \kappa} R_{\sigma \kappa \nu \rho} R_\mu^\rho \nonumber +3 R R^{\mu \nu \sigma \kappa} R_{\mu \nu \sigma \kappa}+
		\\&24 R^{\mu \nu \sigma \kappa} R_{\sigma \mu} R_{\kappa \nu}+16 R^{\mu \nu} R_{\nu \sigma} R_\mu^\sigma-12 R R^{\mu \nu} R_{\mu \nu}).
	\end{align}

Form Eqs.\eqref{L} and \eqref{N},  it is known that the $n$-order lagrangian $\mathcal L$ depends on the different dimensions $d$.
	\begin{itemize}
		\item When $d=4$, the order $n$ is 1. The 1-order lagrangian contains $\mathcal{L}_0$ and $\mathcal{L}_1$ and it is also called the Einstein-Hilbert lagrangian in 4 dimensions ($\mathcal L_0$ and $\mathcal L_1$ are the cosmological constant term and the Einstein term, respectively).
	\item When $d=5$ and $d=6$, the order $n$ is 2. The 2-order lagrangian includes $\mathcal{L}_0$, $\mathcal{L}_1$ and $\mathcal{L}_2$ and it is also called the Einstein-Gauss-Bonnet lagrangian ($\mathcal L_2$ is the Gauss-Bonnet term)~\cite{Glavan2019inb,Garraffo2008hu}.	
	\item By analogy, for $n=3$, the 3-order Lovelock lagrangian contains $\mathcal{L}_0$, $\mathcal{L}_1$ , $\mathcal{L}_2$ and $\mathcal{L}_3$  and it exists in 7 and 8 dimensions ($\mathcal L_3$ is the third-order Lovelock term).
	\item The contribution of higher-order Lovelock terms gets smaller and smaller to the point where it can be ignored. Hence the $n (n \ge 4)$-order Lagrangian can be approximated as the one with the order of 3, and then the 3-order Lovelock theory is used to study black holes in $d\geq 7$ dimensions naturally.
\end{itemize}

Hence, the geometric action of the third-order Lovelock black hole is written as~\cite{Xu2014tja,Xu2016tja},
	\begin{align}
		\mathcal{I} =\dfrac{1}{16\pi G}\int d^dx(\dfrac{\alpha_0}{\lambda^2}\mathcal{L}_0 +\alpha_1\mathcal{L}_1+\alpha_2\lambda^2\mathcal L_2+\alpha_3\lambda^4\mathcal L_3)
		=\dfrac{1}{16\pi G}\int d^dx(R-2\Lambda+\hat\alpha_2\mathcal L_2+\hat\alpha_3\mathcal L_3),
	\end{align}
	here we have taken the liberty of making $\alpha_0=-2\Lambda \lambda^2,\alpha_1=1,\hat\alpha_2=\alpha_2\lambda^2$, and $\hat\alpha_3=\alpha_3\lambda^3$. In the following formulation, we choose $\alpha$ instead of  $\hat\alpha_2$ and $\hat\alpha_3$,
\begin{align}
		\hat\alpha_{2}=\frac{\alpha}{(d-3)(d-4)}, \quad
		\hat\alpha_{3}=\frac{\alpha^2}{72{d-3\choose 4}}.
\end{align}

The static spherical symmetry metric for $d \ge 7$ is expressed as

\begin{equation}
ds^2=-V(r)dt^2+\dfrac{1}{V(r)}dr^2+r^2d\Omega_k^2,
\end{equation}
\begin{equation}
V(r)=k+\dfrac{r^2}{\alpha}\Bigl[1-\Bigl(1+\dfrac{6\Lambda \alpha}{(d-1)(d-2)}+\dfrac{3\alpha m}{r^{d-1}}\Bigr)^{\frac{1}{3}}\Bigr],
\end{equation}
here $m$ is a parameter related to the mass of a black hole, and $k$ is topology of the spacetime curvature and can take $-1$, $0$, and $1$.

The Hawking temperature of the third order black hole in terms of the radius of the event horizon $r_h$ is
\begin{align}
T_h=\dfrac{1}{12\pi r_h(r_h^2+k\alpha)^2}\left[\dfrac{48\pi P r_h^6}{(d-2)}+3(d-3)kr_h^4+3(d-5)\alpha k^2r_h^2+(d-7)\alpha ^2k\right],
\label{T_h}	
\end{align}
where $P$ is pressure via $P=-\Lambda/(8\pi G)$. The entropy conjugated to the temperature reads as
\begin{align}	
S=\dfrac{\sum_kr_h^{d-2}}{4}\left[1+\dfrac{2(d-2)k\alpha}{(d-4)r_h^2}+\dfrac{(d-2)k^2\alpha^2}{(d-6)r_h^4}\right],\  \label{S}
\end{align}
where $\sum_k$ is the volume of the $(d-2)$-dimensional submanifold. Therefore, the thermal potential of the third order Lovelock black hole is expressed as
 \begin{align}
U=&\int(T_h-T)dS\nonumber\\
=&\dfrac{\sum_k r_h^{d-7}}{48 \pi (d-1)}\bigg[48 \pi P r_h^6+(d-1)(d-2)\Big(3kr_h^4+3\alpha k^2r_h^2+\alpha^2k\Big)\bigg]\nonumber\\
&-\dfrac{\sum_kr^{d-2} }{4} \bigg[1+\dfrac{2(d-2)k\alpha}{(d-4)r_h^2}+\dfrac{(d-2)k^2\alpha^2}{(d-6)r_h^4}\bigg]T.
\label{U}
 \end{align}

Now we place various thermal states in the black hole thermodynamic system at the extreme points of the potential function and it meets the expressions
\begin{equation}
	f(r_h)=\dfrac{dU(r_h)}{dS(r_h)}=0.
	\label{f}
\end{equation}
Then we can turn the thermodynamic problems into solving the zeroes of the real function $f(r_h)$. To see the full picture of the problem, we need to change the real function $f(r_h)$ into a complex continuation
function $f(z)$, and use the method of complex analysis to study~\cite{Xu2023vyj}.

In complex analysis, the $Argument ~ Principle$ is an effective method to calculate the number of zeros of analytic functions. If $f(z)$ is a meromorphic function in a simple closed contour  $C$, and is analytically nonzero on  $C$, then there is:
\begin{equation}
	N(f, C)-P(f, C)=\frac{1}{2 \pi i} \oint_C \frac{f^{\prime}(z)}{f(z)} \mathrm{d} z=\frac{\Delta_C \arg f(z)}{2 \pi},
	\label{Argument}
\end{equation}
where $N(f, C)$ and  $P(f, C)$ are respectively the number of zeros and poles of $f(z)$ in  $C$, $f'(z)$ is the first order derivative of $f(z)$, and $\text{arg}f(z)$ is the argument of $f(z)$.
Making a transformation $\omega = f(z)$, the above equation is then expressed as the number of rotations of $\omega$ around the origin of curve $C'$ as the complex variable $z$ moves around the complex envelope $C$, where $C'$ is the image curve of $C$ after the transformation. The winding number is denoted by
\begin{equation}
	W:=\frac{1}{2 \pi i} \oint_{C'} \frac{\mathrm{d} \omega}{\omega}=\frac{1}{2 \pi i} \oint_C \frac{f^{\prime}(z)}{f(z)} \mathrm{d} z.
	\label{W}
\end{equation}
 If the analytic function $f(z)$ does not have poles within the complex perimeter, then the winding number of about the origin is $W=N(f, C)$.
When the complex variable $z$ varies on contour $C$,  the image of the argument function $\theta = \text{arg}f(z)$ can be a Riemann surface. The winding number of the origin corresponds to the foliations of the Riemann surface of the complex variable function.

When $W=1$, there is no phase transition, when $W=2$, it corresponds to the second-order phase transition, and when $W=3$, it means that the first-order phase transition will occur, accompanied by the second-order phase transition~\cite{Xu2023vyj}. Here, we consider only zeros that are real and positive, and these correspond to physical values for the radius $r_h$. Next we hope to use this method to predict the structure of phase transitions of third-order Lovelock black hole. For the case $k = 0$ of the topology of the spacetime curvature, the phenomena are the same as an ideal gas and have no phase changes. Therefore we will focus on the two cases $k = -1$ and $k = +1$.
	
\section{Hyperbolic topology}\label{hyper}
\begin{figure}[htb]
	\begin{center}
		\includegraphics[width=50 mm]{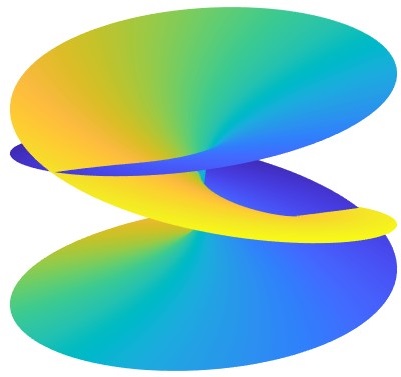}
	\end{center}
	\caption{Riemann surface of the first order and second order phase transitions for the black hole system.}
	\label{fig20}
\end{figure}

In this case, we have $k=-1$. For the Lovelock black hole in the hyperbolic case, the analytic function $f(z)$ is calculated by Eqs.~\eqref{U} and~\eqref{f} as follows
\begin{equation}
	f(z)=\dfrac{1}{12\pi z(z^2-\alpha)^2}\bigg[\dfrac{48\pi P z^6}{(d-2)}-3(d-3)z^4+3(d-5)\alpha z^2-(d-7)\alpha ^2-12\pi z T(z^2-\alpha)^2\bigg].
\end{equation}
Whether $d=7$ or either $7<d\le 12$, this analytic function has three zeros at most in the entire complex plane $\mathbf{C}$ with the singularities removed. The only difference between them is that the singularities are $\pm\sqrt{\alpha}$ for $d=7$, whereas for $7<d\le 12$, the singularities are $0$ and $\pm\sqrt{\alpha}$. Hence we obtain the winding number $W=3$ and its complex structure is the Riemann surface with three foliations, as shown in Fig.~\ref{fig20}. Based on the results of the study~\cite{Xu2023vyj}, we predict that the black hole will undergo the phase transitions with first and second orders.

\begin{figure}[htb]
	\centering
	\subfigure[$p<p_m$]{
		\includegraphics[width=5cm]{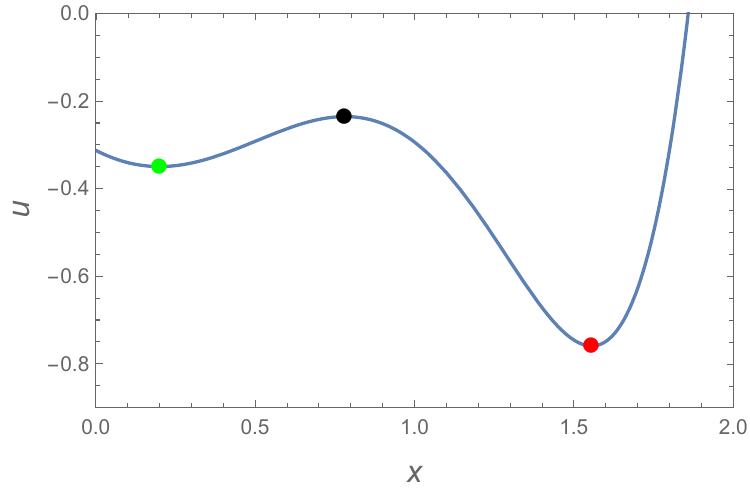}}
	\subfigure[$p=p_m$]{
		\includegraphics[width=5cm]{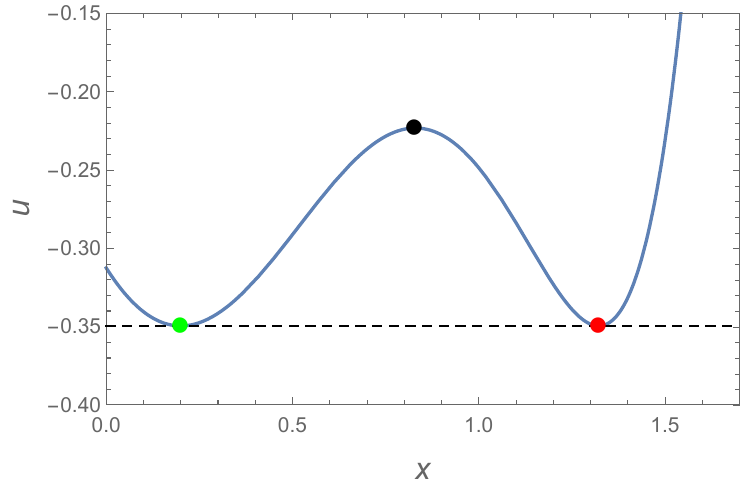}}
	\subfigure[$p>p_m$]{
		\includegraphics[width=5cm]{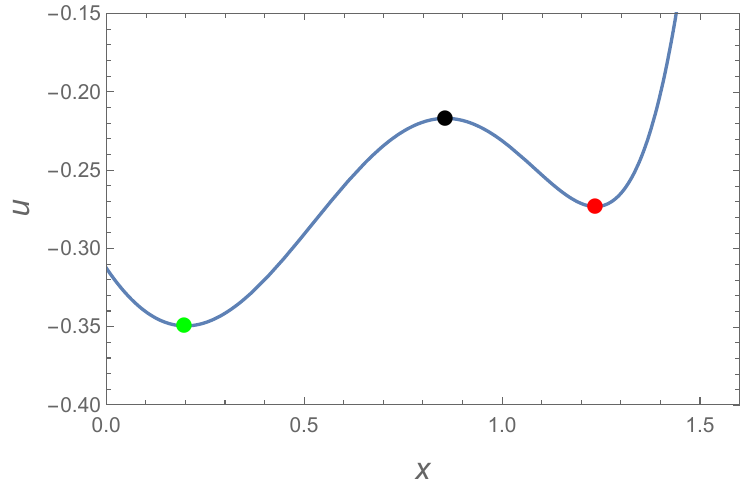}
		
		\includegraphics[width=5cm]{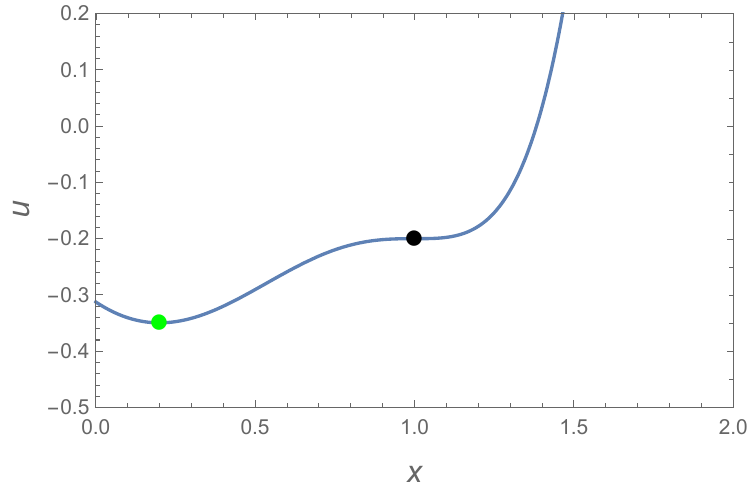}
		
		\includegraphics[width=5cm]{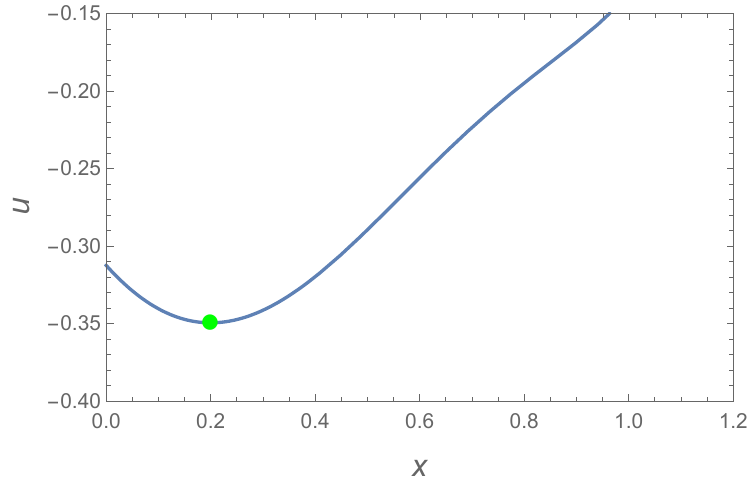}}
	\caption{The $u-x$ plots of $t=0.2$ for $d = 7$. The \textcolor[rgb]{0.00,1.00,0.00}{$\bullet$}-phase in the diagram represents the small black hole state, the \textcolor[rgb]{1.00,0.00,0.00}{$\bullet$}-phase represents the large black hole state and the {$\bullet$}-phase represents the unstable black hole state.  The pressure $p$ increases from left to right in $p>p_m$ plots.}
	\label{fig2}
\end{figure}

Since $d=7$ is of the same type as $d>7$, let's make a long story short and just use $d=7$ as an example to verify the above viewpoint. We easily know that there is only one set of critical points in the hyperbolic case, and obtain the critical points for $d = 7$ from~\cite{Xu2014tja, Xu2016tja}
\begin{equation}
	P_c=\dfrac{5}{8\pi \alpha},\qquad T_c=\dfrac{1}{2\pi \sqrt{\alpha}},\qquad v_c=\dfrac{4}{5}\sqrt{\alpha},
	\label{cp1}
\end{equation}
where $v=4r_h/(d-2)$. For the sake of discussion, we introduce the following dimensionless thermodynamic quantities
\begin{equation}
	p:=\dfrac{P}{P_c},\qquad t:=\dfrac{T}{T_c},\qquad x:=\dfrac{r_h}{r_c},\qquad t_h:=\dfrac{T_h}{T_c},\qquad s:=\dfrac{S}{S_c},\qquad u:=\dfrac{U}{|U_c|}.
	\label{dimensionless}
\end{equation}
The validity of the method is now checked with an analysis of the behavior of the thermal potential. After a series of calculations, we obtain the dimensionless thermal potential for $d = 7$
\begin{equation}
	\begin{split}
	u&=\dfrac{5}{16}(px^6-3x^4+3x^2-1)-\dfrac{3}{8}t(x^5-\dfrac{10}{3}x^3+5x).
	\label{U1}
	\end{split}
\end{equation}
From Eq.~\eqref{U1} it can be seen that the two key parameters ($p$ and $t$) affect the behaviors of the thermal potential.  Here we fix the parameter $t$ to observe the variation of the thermal potential with $p$. In Fig.~\ref{fig2} we show  the $u-x$ plot at $d = 7$.
According to Eqs.~\eqref{dU} and~\eqref{variation}, we know that the black hole state can be placed at an extreme value of the thermal potential. An unstable black hole state is at the local maximum point, and a stable black hole state is at the minimal point. The lower the potential, the higher the probability that the black hole is at that point and the more stable the system is.

From the diagrams (a) and (b) in Fig.~\ref{fig2}, we obtain that at a fixed temperature $t=0.2$ (for any value $ 0 < t < 1$, we always get the same result), a global minimum and a local minimum start to change as the pressure $p$ increases from $p=0$ to $p=p_m$ at which the global minima of the thermal potential are equivalent. Specifically, at $0 < p < p_m$, the thermal potential of the large black hole phase is lower than that of the small black hole phase, implying that the  system tends towards the large black hole phase. When $p$ increases to $p_m$, the large black hole phase and the small black hole phase are in equilibrium.
Similarly, it is clear from the diagrams (b) and (c) that as $p$ increases, the two equivalent global minimums begin to change. Small black hole phase is at the global minimum, while the large black hole phase changes to be in the local minimum until it disappears. Specifically, the thermal potential of the small black hole phase is lower than that of the large black hole phase, which means that the system tends to the small black hole phase at $p>p_m$.

Thus, it is clear from the above analysis that in the $k=-1$ hyperbolic case, the system has a first-order phase transition from a large black hole to a small black hole. From Eq.~\eqref{cp1}, it follows that there is a critical point, which ia the inflection point of the curve, therefore the system also has a second-order phase transition. This is exactly what we predicted.

\section{Spherical topology}\label{spher}
In this case, we have $k=+1$. For the Lovelock black hole in the spherical case, the analytic function is calculated by the Eqs.~\eqref{U} and~\eqref{f},
\begin{equation}
	f(z)=\dfrac{1}{12\pi z(z^2+\alpha)^2}\bigg[\dfrac{48\pi P z^6}{(d-2)}+3(d-3)z^4+3(d-5)\alpha z^2+(d-7)\alpha ^2-12\pi z T(z^2+\alpha)^2\bigg].
	\label{f1}
\end{equation}

Here we note that the zeroes of the cases in $d = 7$ and $d>7$ are not equal across the complex plane with all singularities removed, which leads to different winding number and Riemann surfaces. So the spherical case is not as straightforward as the hyperbolic one and needs to be discussed differently.

In particular at $d=12$ there is only one zero point, so the winding number is $1$ and it is a single-foliation Riemann surface, resulting that the system does not undergo a phase transition. This conclusion is already well-known and is not elaborated here.

\subsection{$d=7$}

For $d=7$,  we can obtain its analytic function from Eq.~\eqref{f1} expressed as
\begin{equation}
	f(z)=\dfrac{1}{10\pi (z^2+\alpha)^2}\bigg[(8P\pi z^5+10z^3+5\alpha z)-10\pi T (z^2+\alpha)^2\bigg].
\end{equation}
Similarly, there are three zeroes at most on the complex plane $\mathbf{C}\backslash\{ \pm \sqrt{\alpha}i\}$, so that the winding number $W=3$ and the complex structure is similar to the hyperbolic case in $d=7$ with three foliations. Hence, we predict that there will have second-order and first-order phase transitions. Next, we verify its correctness by  the thermal potential.

Firstly from~\cite{Xu2014tja} we obtain the critical points
\begin{equation}
	P_{c1}=0,\qquad T_{c1}=0,\qquad v_{c1}=0,
\end{equation}
and
\begin{equation}
	P_{c2}=\dfrac{17}{200\pi \alpha},\qquad T_{c2}=\dfrac{1}{\pi \sqrt{5\alpha}},\qquad v_{c2}=\dfrac{4}{5}\sqrt{5\alpha}.\label{cp2}
\end{equation}
Then by use of the Eqs.~\eqref{U} and~\eqref{dimensionless}, we can obtain the expression for the dimensionless thermal potential in $d = 7$,
\begin{equation}
	\begin{split}
		u&=\dfrac{1}{4}(17px^6+75x^4+15x^2+1)-t(15x^5+10x^3+3x).
	\end{split}
	\label{U2}
\end{equation}

\begin{figure}[htb]
	\centering
	
	\subfigure[$p<p_m$]{
		\includegraphics[width=5cm]{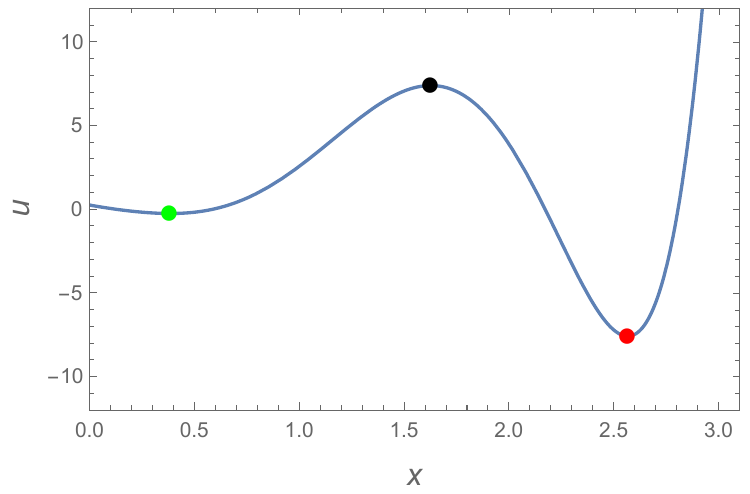}}
	\subfigure[$p=p_m$]{
		\includegraphics[width=5cm]{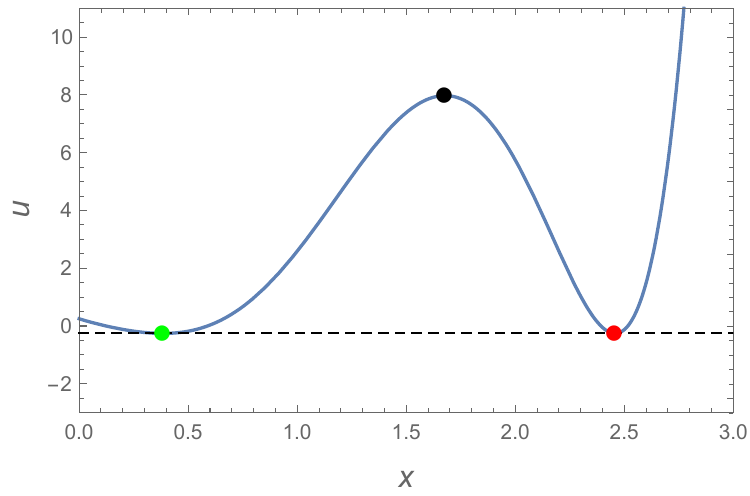}}
	\subfigure[$p>p_m$]{
		\includegraphics[width=5cm]{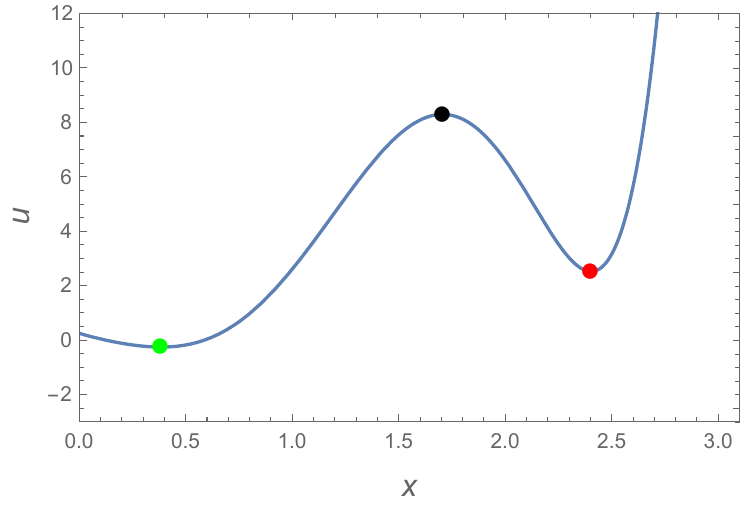}
		
		\includegraphics[width=5cm]{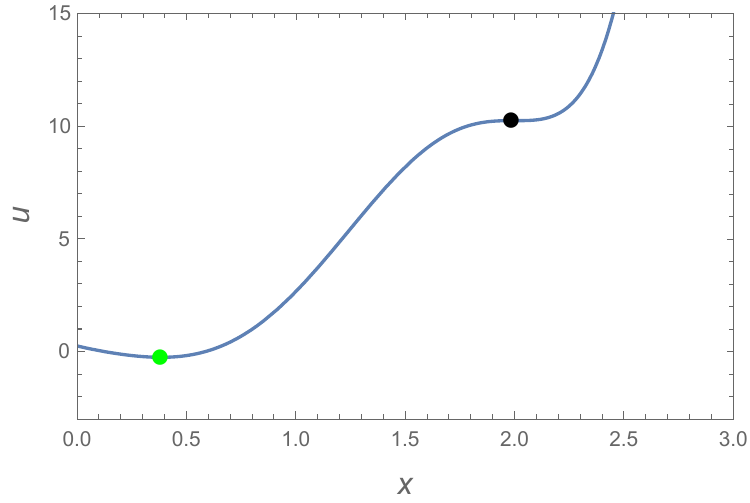}
		
		\includegraphics[width=5cm]{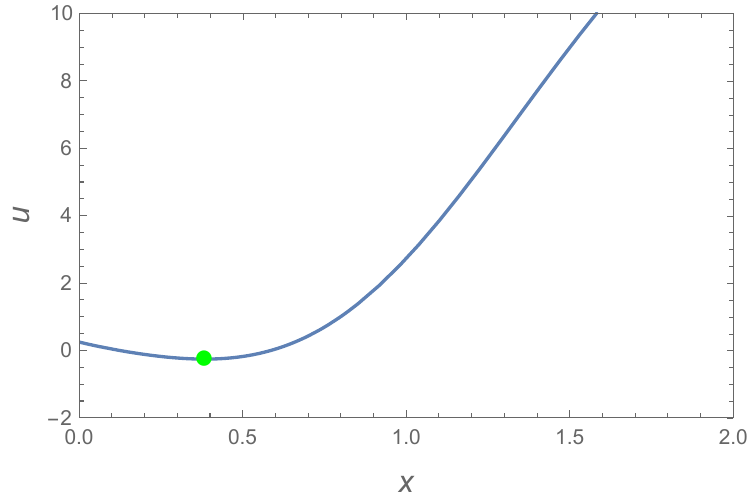}}
	\caption{The $u-x$ plots of $t=0.8$ for $d = 7$. The \textcolor[rgb]{0.00,1.00,0.00}{$\bullet$}-phase in the diagram represents the small black hole state, the \textcolor[rgb]{1.00,0.00,0.00}{$\bullet$}-phase represents the large black hole state and the {$\bullet$}-phase represents the unstable black hole state. The pressure $p$ increases from left to right in $p>p_m$ plots.}
	\label{fig3}
	\end{figure}

Surprisingly, its behaviors are extremely similar to those in the hyperbolic case. From the diagrams (a) and (b) in Fig.~\ref{fig3}, we get that at a fixed temperature $t=0.8$ (we always get the same results when taking any value $0 < t < 1$) when the pressure $p$ starts increasing from $0$ to $p_m$, a global minimum and a local minimum start to become two equivalent global minimum values of thermal potential. Specifically, at first the thermal potential of the large black hole phase is lower than that of the small black hole phase, which means that the black hole system tends towards the large black hole phase. Gradually there is a clear upward trend in the large black hole phase, and finally the large black hole phase is at the same level as the small black hole phase. From the diagrams (b) and (c), as the pressure increases from $p_m$, the two equivalent global minima start to change, with the small black hole phase being a global minimum and the large black hole phase becoming a local minimum until it disappears. This means that the black hole system tends to the small black hole phase at $p>p_m$.

When $p<p_m$, the whole black hole system will be completely in the large black hole phase, and conversely, the system is completely in the small black hole phase at $p>p_m$. There is also a critical points Eq.~\eqref{cp2} under this dimensions. So it is concluded that, the system will have first-order and  second-order phase transitions. This is the same result as calculated by the winding number.

\subsection{$d>7$}

Let now study the cases of $8$, $9$, $10$, and $11$ dimensions. From Eq.~\eqref{f1}, it follows that the $8$, $9$, $10$, and $11$-dimensional cases are similar.  Therefore, we take the case of $d=9$ as an example.
The analytic function is obtained by substituting $d=9$ into Eq.~\eqref{f1}, which reads as
\begin{equation}
	f(z)=\dfrac{1}{42\pi z(z^2+\alpha)^2}\bigg[24P\pi z^6+63z^4+42\alpha z^2+7\alpha^2-42\pi zT( z^2+\alpha)^2\bigg]
\end{equation}
There are four zeroes at most on the complex plane $\mathbf{C}\backslash\{\pm \sqrt{\alpha}i, 0\}$. Hence, the winding number is $W=4$ and the complex structure is the Riemann surface with four foliations.

\begin{figure}[htb]
 	\begin{center}
 		\includegraphics[width=130 mm]{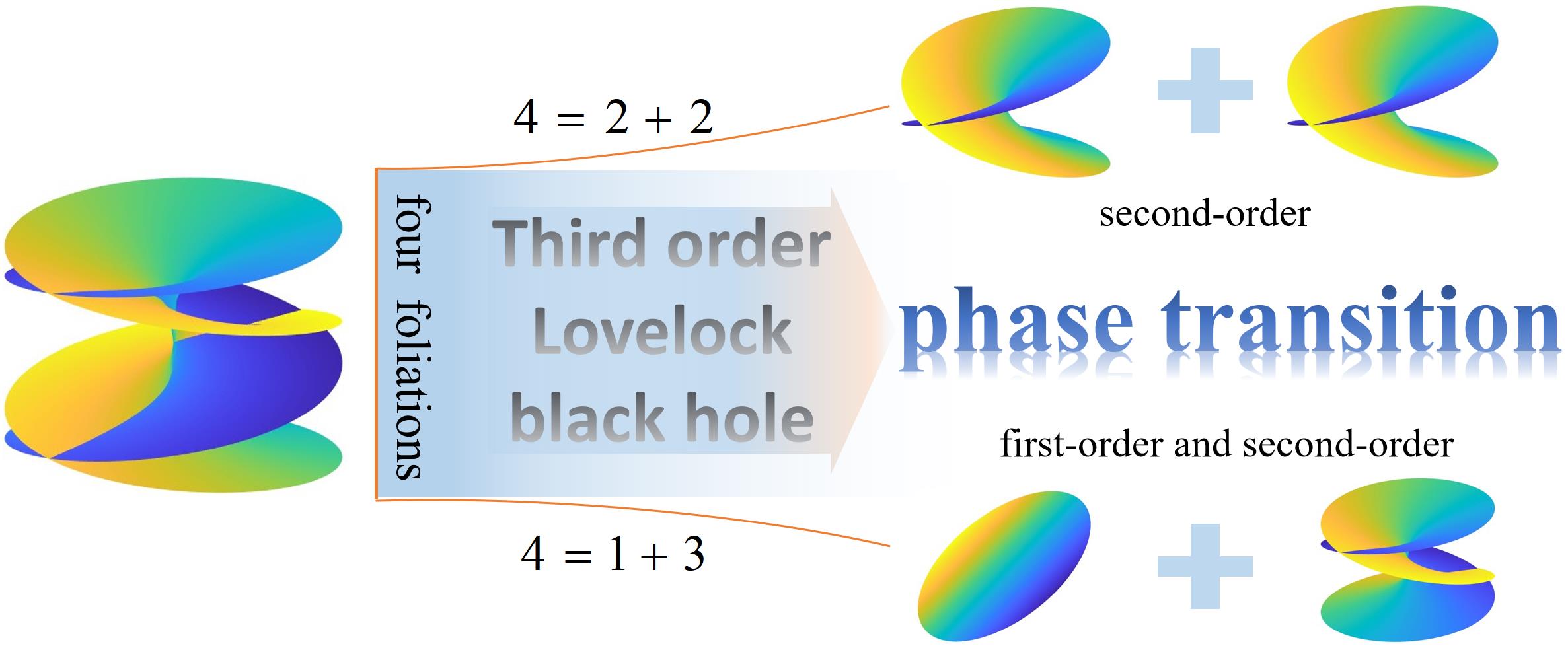}
 	\end{center}
 	\caption{Two decompositions with a winding number $W=4$.}
 	\label{fig22}
 \end{figure}

According to the basic elements that the corresponding relationship between winding number and phase transition, we can find that $W=4$ can be decomposed in two ways:
(i) $4=2+2$, it means that the system only has two second-order phase transitions; (ii) $4 =1+3$,  it shows that the system will have one first-order and one second-order phase transitions. Of course, there is a clearer breakdown in Fig.~\ref{fig22}. So we conjecture that in $d=9$ there will be two different types of phase transition.

For the case of $d > 7$, there are two pairs of critical points for the system. This makes dimensionless processing more complicated, so we do not do this here, which is slightly different from the previous analysis. The two sets of critical points in $d=9$ were obtained from~\cite{Xu2014tja} which can be written as
\begin{equation}
P_{c1}=\frac{63}{16 \pi \alpha(6-\sqrt{21})^3(\sqrt{21}-21)},\quad T_{c1}=\frac{3\sqrt{3}( \sqrt{21}-7)}{\sqrt{\alpha}\pi \sqrt{6-\sqrt{21}} (\sqrt{21} -21)}, \quad v_{c1}=\frac{4 \sqrt{\alpha} \sqrt{18-3 \sqrt{21}}}{21},
\label{cp3}
\end{equation}
and
\begin{equation}
P_{c2}=\frac{63}{16 \pi \alpha(6+\sqrt{21})^3(\sqrt{21}+21)},\quad	T_{c2}=\frac{3\sqrt{3}( \sqrt{21}+7)}{\sqrt{\alpha}\pi \sqrt{6+\sqrt{21}} (\sqrt{21} +21)}, \quad v_{c2}=\frac{4 \sqrt{\alpha} \sqrt{18+3 \sqrt{21}}}{21}.
\label{cp4}
\end{equation}
The thermal potential is expressed with the help of Eq.~\eqref{U} as
\begin{equation}
U=\frac{\sum_k}{4}\left[\frac{7}{12\pi}\left(\frac{6}{7}\pi P r_h^8+3r_h^6+3\alpha r_h^4+\alpha^2r_h^2\right)-T\left(r_h^7+\frac{14}{5}\alpha r_h^5+\frac{7}{3}\alpha^2r_h^3\right)\right].
	\label{U3}
\end{equation}
For the sake of simplicity, we set both $\alpha=1$ and $\sum_k=1$. By analysis, we find that the phase transition between the two critical temperatures needs to be discussed on a case-by-case basis.

\paragraph[(a)]{$T_{c1}\le T< T_{cm}$}

\begin{figure}[htb]
	\centering
	
	\subfigure[]{
		\includegraphics[width=5.2cm]{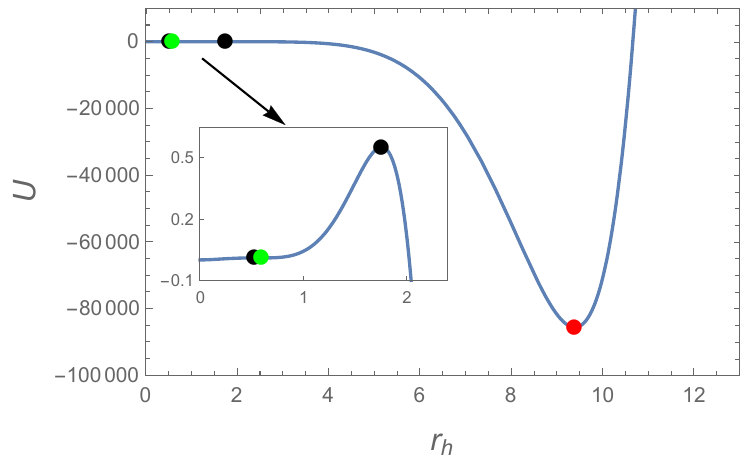}}
	\subfigure[]{
		\includegraphics[width=5cm]{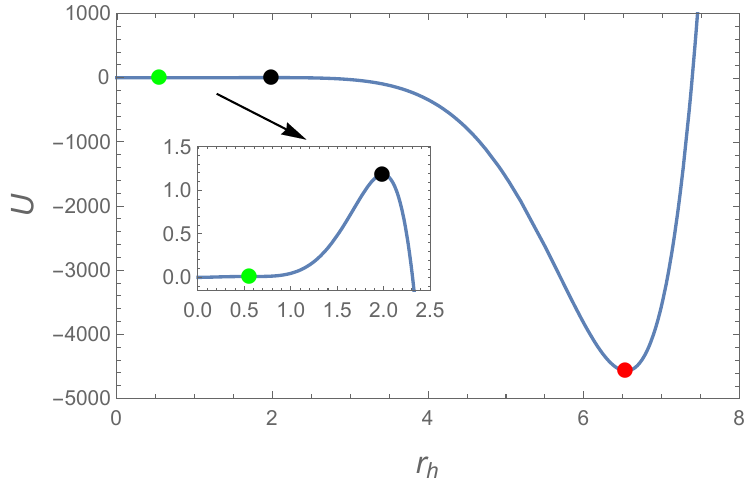}}
	\subfigure[]{
		\includegraphics[width=4.8cm]{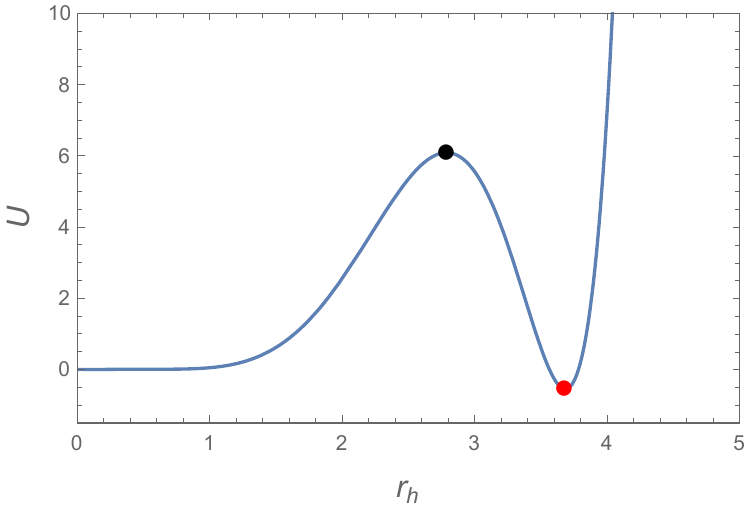}}
	\subfigure[]{
		\includegraphics[width=5cm]{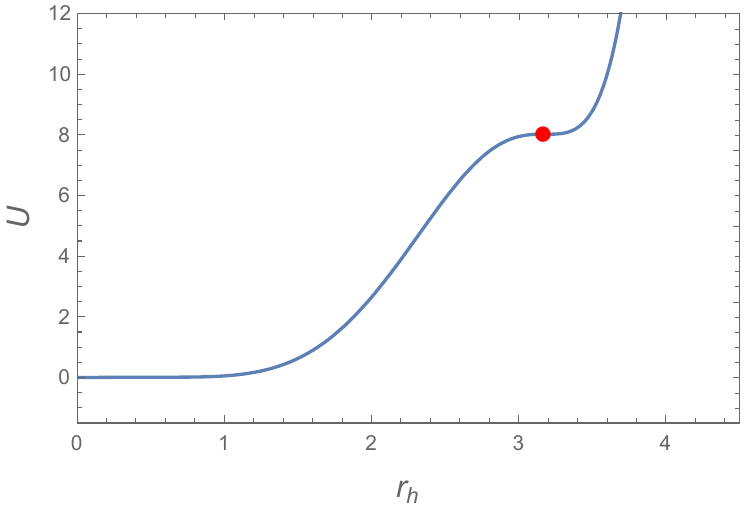}}
	\subfigure[]{
		\includegraphics[width=5cm]{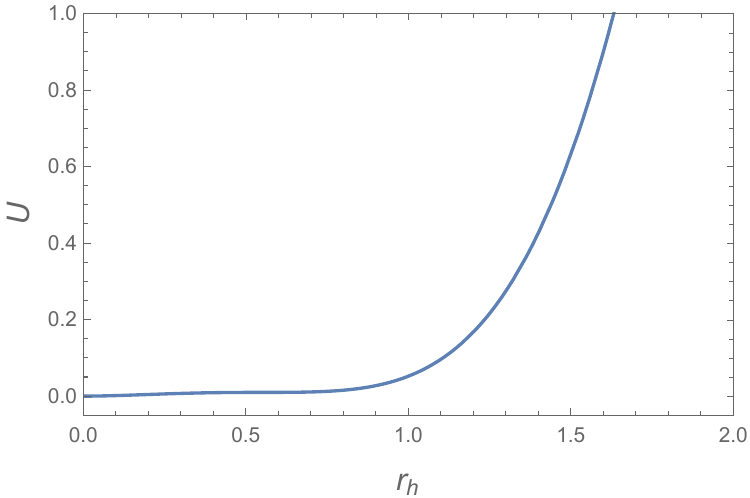}}
	\caption{The $U-r_h$ plots of $T=0.2075$ for  $d=9$. The \textcolor[rgb]{0.00,1.00,0.00}{$\bullet$}-phase in the diagram represents the small black hole state, the \textcolor[rgb]{1.00,0.00,0.00}{$\bullet$}-phase represents the large black hole state and the {$\bullet$}-phase represents the unstable black hole state. The pressure $P$ increases from the diagrams (a) to (e).}
	\label{fig4}
\end{figure}

We can see that from the Fig.~\ref{fig4}, there is a gradual merging between the extremal points until they disappear as the pressure $P$ increases, while during this time the large black hole phase is always a global minimum and there is no transition between two minima. This means that the system is no first-order phase transition. Instead, the system has a second-order phase transition due to the presence of the inflection point Eq.~\eqref{cp3}.

\paragraph[(b)]{$T=T_{cm}$}

\begin{figure}[htb]
	\centering
	\subfigure[$P<P_m$]{
		\includegraphics[width=5cm]{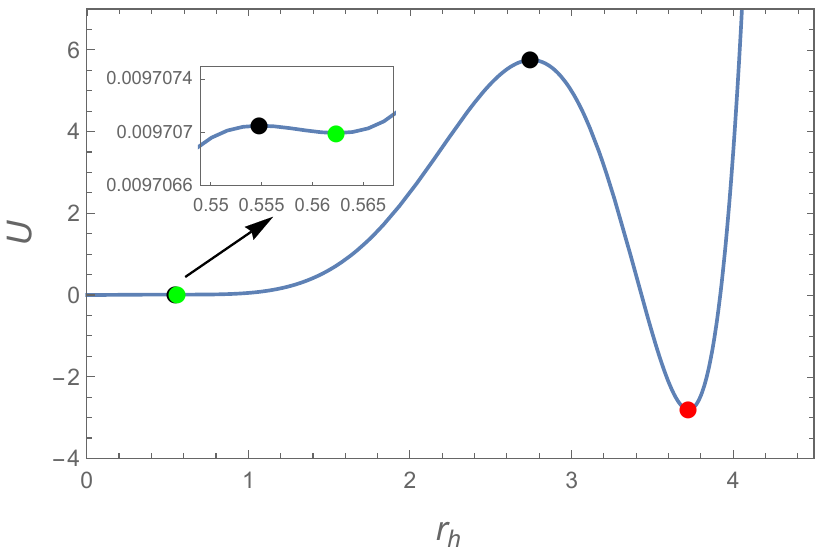}}
	\subfigure[$P=P_m$]{
		\includegraphics[width=5cm]{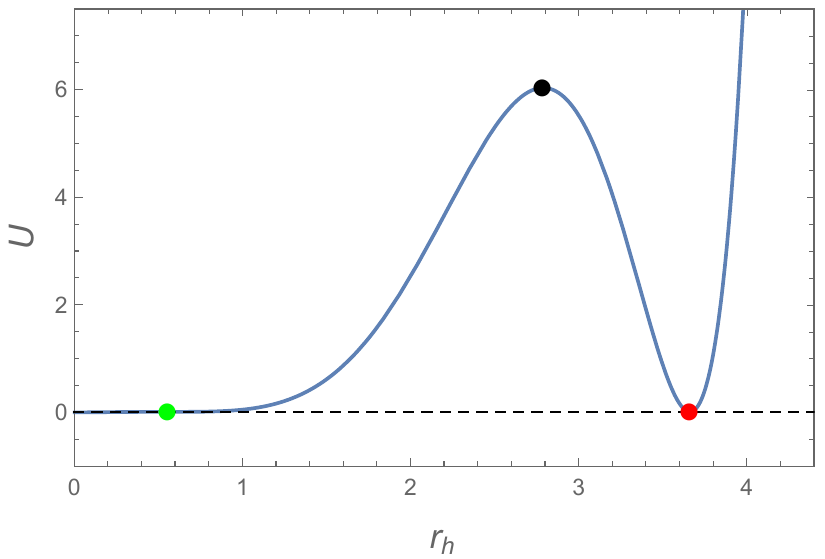}}
	
	\subfigure[$P>P_m$]{
		\includegraphics[width=5cm]{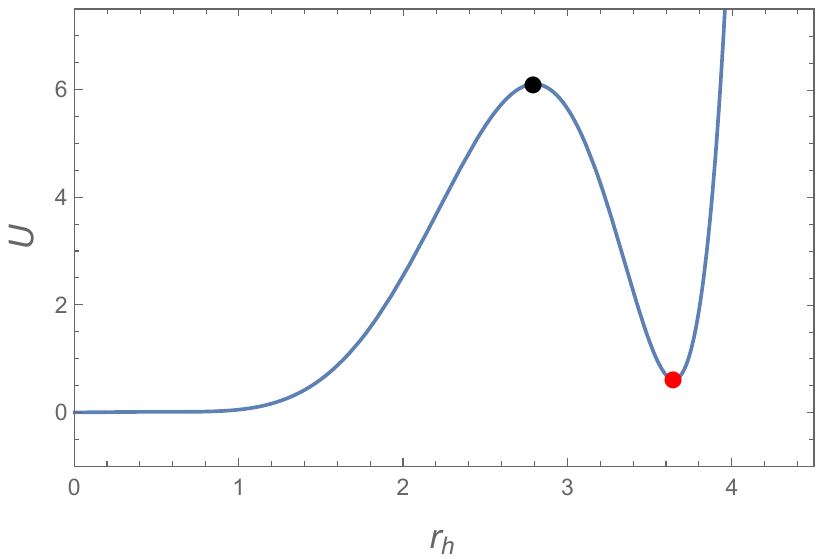}\qquad
		
		\includegraphics[width=5cm]{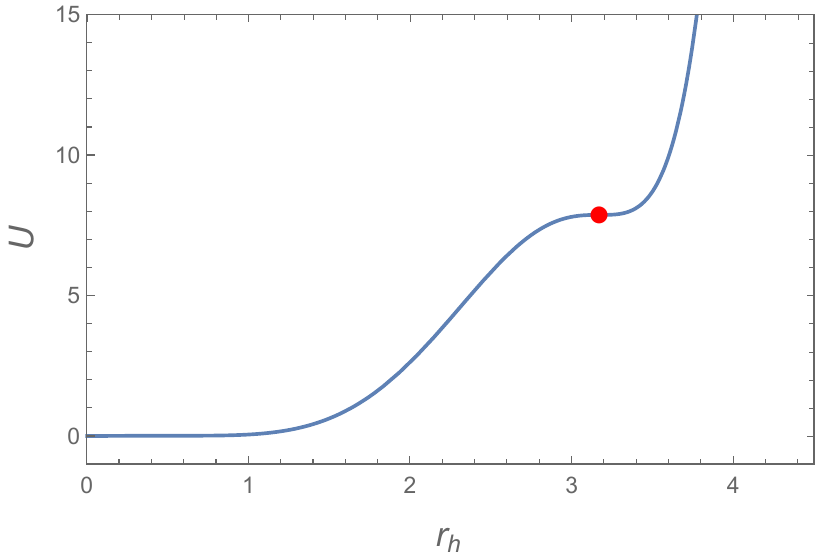}\qquad
		
		\includegraphics[width=5cm]{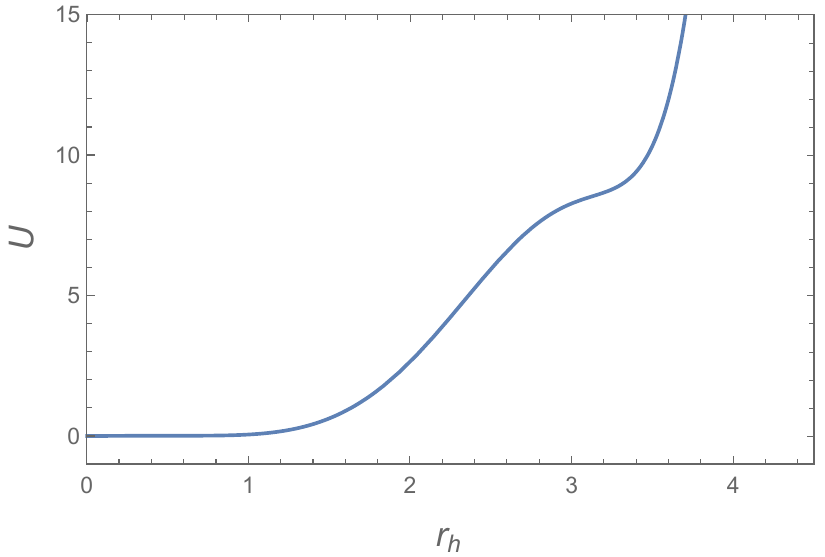}	}
	\caption{The $U-r_h$ plots of $T=T_{cm}$  for $d=9$. The \textcolor[rgb]{0.00,1.00,0.00}{$\bullet$}-phase in the diagram represents the small black hole state, the \textcolor[rgb]{1.00,0.00,0.00}{$\bullet$}-phase represents the large black hole state and the {$\bullet$}-phase represents the unstable black hole state. The pressure $P$ increases from left to right in $P>P_m$ plots.}
	\label{fig5}
\end{figure}

From the Fig.~\ref{fig5}, we can see that the thermal potential changes similarly to that of $T<T_{cm}$ at both $P<P_m$ and $P>P_m$. While it is worth noting that at $P=P_m$, the global minimum and the local minimum become two equal global minima, a phenomenon that does not exist for $T<T_{cm}$. Therefore,  $T_{cm}$ is the point at which the phase transition will  begin to occur, which is still a second-order phase transition.

\paragraph[(c)]{$T_{cm}<T\le T_{c2}$}

\begin{figure}[htb]
	\centering
	\subfigure[$P<P_m$]{
		\includegraphics[width=5cm]{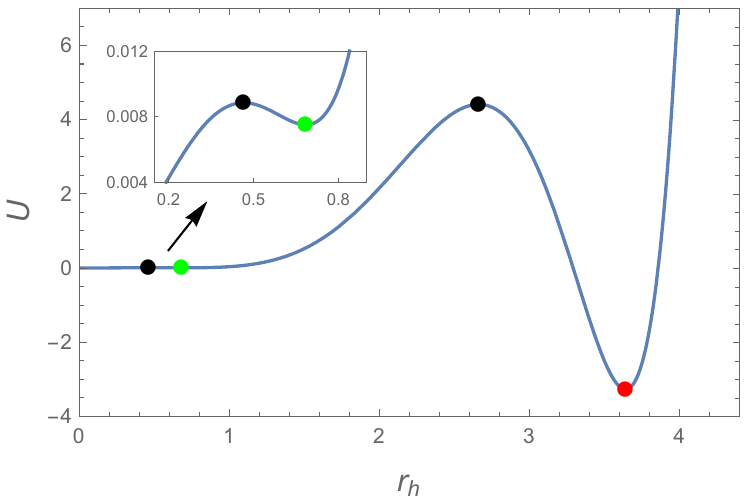}}
	\subfigure[$P=P_m$]{
		\includegraphics[width=5cm]{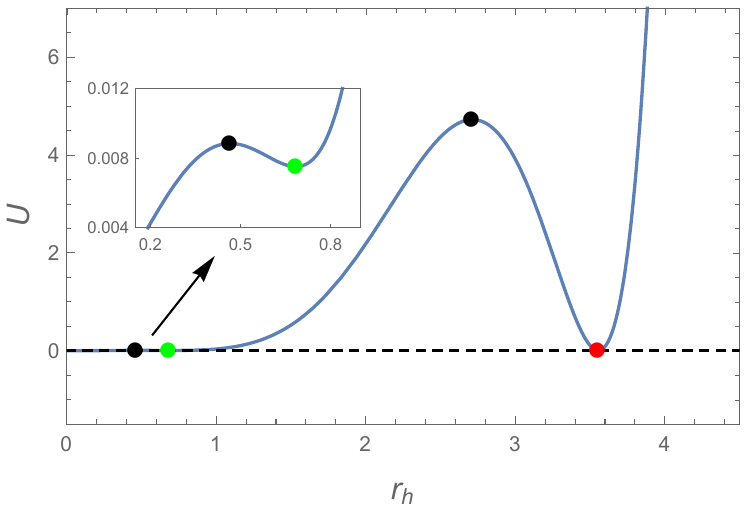}}
	
	\subfigure[$P>P_m$]{
		\includegraphics[width=5cm]{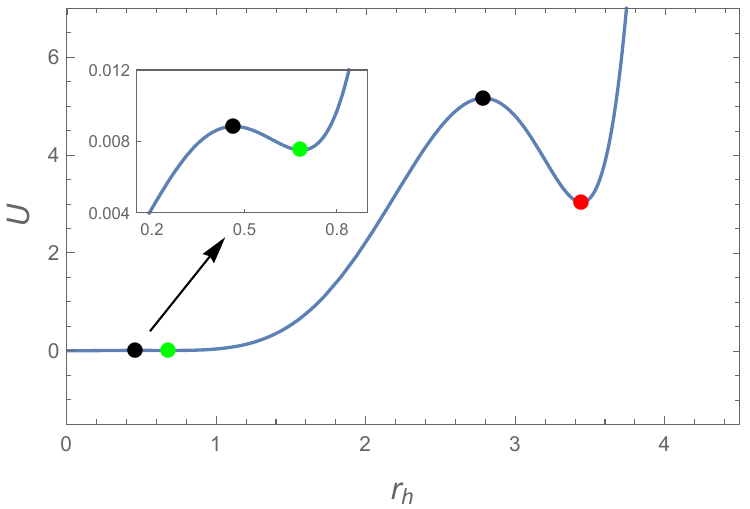}  \qquad
		
		\includegraphics[width=5cm]{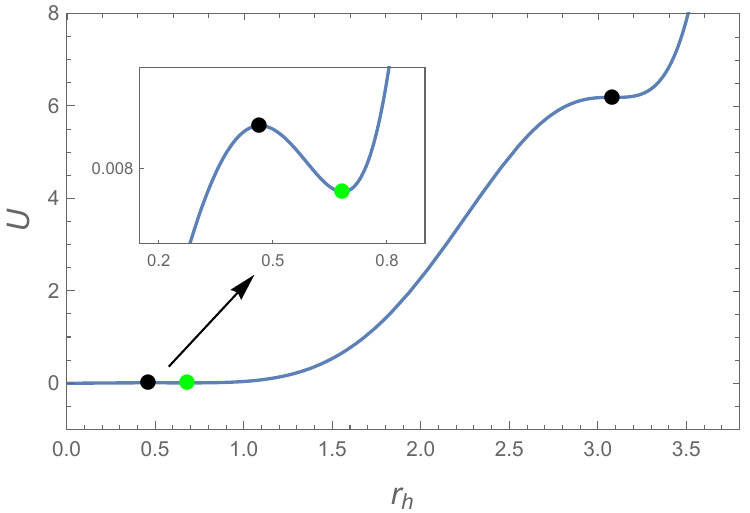} \qquad
		
		\includegraphics[width=5.2cm]{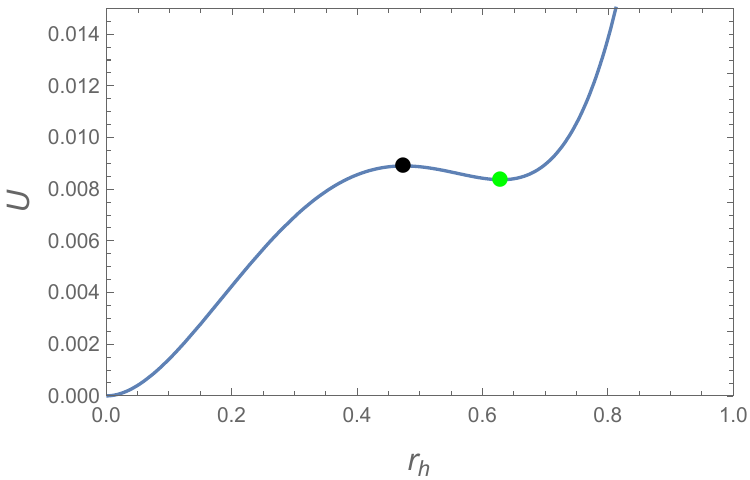}	}
	\centering
	\subfigure[$P>P_m$]{
		\includegraphics[width=5cm]{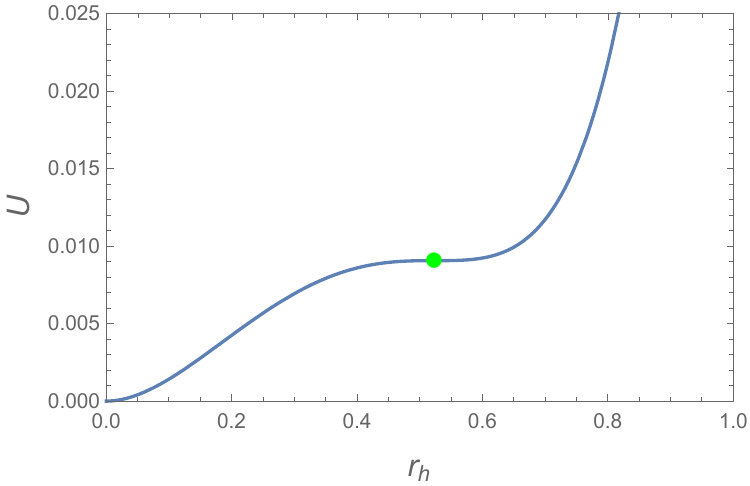}\qquad
		
		\includegraphics[width=5cm]{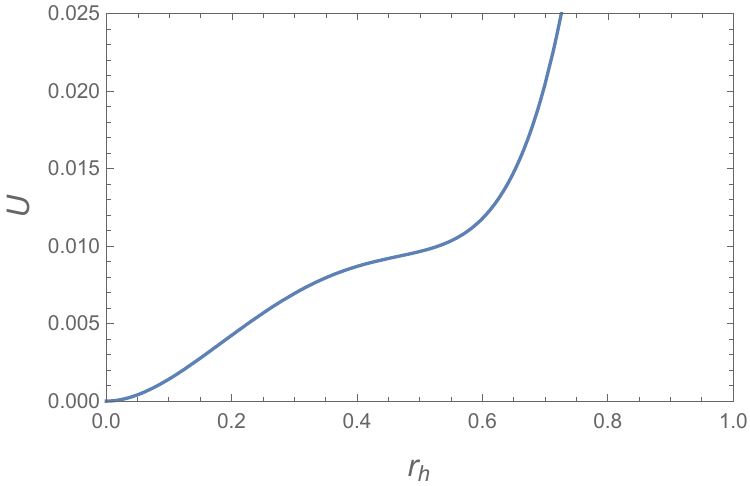}}	
	\caption{The  $U-r_h$ plots of $T=0.21$ for  $d=9$. The \textcolor[rgb]{0.00,1.00,0.00}{$\bullet$}-phase in the diagram represents the small black hole state, the \textcolor[rgb]{1.00,0.00,0.00}{$\bullet$}-phase represents the large black hole state and the {$\bullet$}-phase represents the unstable black hole state. The pressure $p$ increases from left to right in $P>P_m$ plots.}
	\label{fig6}
\end{figure}

From the diagrams (a) to (b) in Fig.~\ref{fig6}, as the pressure $P$ starts to rise, it can be seen that the global minimum of the large black hole phase and the local minimum of the small black hole phase change to two equivalent global minima.  The black hole system changes from a large black hole phase to a co-existing large and small black hole phases. From the diagrams (b) to (d), the pressure continues to increase from $P_m$, the two global minima are transformed into a global minima and a local minima ,until the local minima disappear. The thermal potential of the small black hole phase is lower than that of the large black hole phase, which means that the system tends towards the small black hole phase. From the above analysis, it is clear that the system has a first-order phase transition. Meanwhile due to the inflection point Eq.~\eqref{cp4}, it also has a second-order phase transition.

We conclude from the thermal potential diagrams that there are indeed two different phase transition processes in $d = 9$, perfectly verifying the previous conjecture.

\section{Summary}\label{Summary}
In this paper, the complex structure of third-order Lovelock black hole phase transition is predicted by the local winding number, and its accuracy is verified by the behavior of thermal potential. By transposing the complex analysis from mathematics to study the microstructure of the black hole thermodynamics and relating the winding number to the type of phase transition, it is easy to know the order of the different phase transitions of the black hole.

In the hyperbolic case of arbitrary dimensions and the spherical case of $7$ dimensions, the winding number is $W=3$ and the complex structure is the Riemann surface with three foliations, which indicates that there are first-order and second-order phase transitions in this system. The winding number is $W=4$ in $7<d < 12$ for the spherical case and the corresponding complex structure is four-foliations Riemann surface.

The thermal potential is next used to explore specifically how a black hole changes from one state to another. The thermal potential of systems with varying pressure reveals different properties. Based on the nature of the thermal potential, the phase transition processes of Lovelock black holes under different topologies are analysed.

For $k=-1$, the system has first-order and second-order phase transitions.

For $k=+1$, the situation is slightly more complicated. The phase transition process in 7 dimensions  is similar to that of the hyperbolic case, which there are also the first-order and second-order phase transitions. While in $8,9,10,11$ dimensions, there is the key intermediate temperature $T_{cm}$. When the temperature is $T_{c1}<T< T_{cm}$, there are only second-order phase transitions, and when $T_{cm}<T< T_{c2}$, the system has both second-order and first-order phase transitions. As the winding number tells that:

(i)~$4=2+2$ states that only second-order phase transition occurs. It is of this type when the temperature is between $T_{c1}$ and $T_{cm}$ for the Lovelock black holes in the spherical topology of $d>7$ dimensions.

(ii)~$4=1+3$ indicates that the system has both first-order abd second-order phase transitions. It occurs when the temperature is between  $T_{cm}$ and $T_{c2}$ for the Lovelock black holes in the spherical topology of $d>7$ dimensions.

The results of the above thermal potential analysis perfectly match the one of winding number prediction. By establishing the connection between the winding number and the black hole phase transition, we get the complex phase transition structure. Complex analysis is an effective method to further study the microstructure of black hole systems. We hope that this work will provide new ideas for the study of black hole thermodynamic phase transitions, and thus further enrich the content of black hole thermodynamics.

\section*{Acknowledgments}
This research is supported by National Natural Science Foundation of China (Grant No. 12105222, No. 12275216, and No. 12247103).

\end{document}